\documentstyle[twocolumn,pra,aps,epsfig,graphicx]{revtex}

\begin{document}
\draft
\author{C. Figueira de Morisson Faria$^1$, R. Kopold$^2$, W. Becker$^2$\thanks{also at Center for Advanced Studies, Department of Physics and Astronomy, University of New Mexico, Albuquerque, NM 87131}, and J. M. 
Rost$^{1}$}
\address{$^1$Max Planck Institut f\"{u}r Physik komplexer Systeme,
N\"{o}thnitzer Str. 38, D-01187 Dresden, Germany \\
$^2$Max Born Institut f\"{u}r nichtlineare Optik und
Kurzzeitspektroskopie,
Max Born Str. 2A, D-12489 Berlin, Germany}
\title{Resonant enhancements of high-order harmonic generation}
\date{\today}
\maketitle

\begin{abstract}
Solving the one-dimensional time-dependent Schr\"odinger equation for simple model potentials, we investigate resonance-enhanced high-order harmonic generation, with emphasis on the physical mechanism of the enhancement. By truncating a long-range potential, we investigate the significance of the long-range tail, the Rydberg series, and the existence of highly excited states for the enhancements in question. We conclude that the channel closings typical of a short-range or zero-range potential are capable of generating essentially the same effects.
\end{abstract}
\pacs{32.80.Rm, 42.65.Ky}

\section{Introduction}

A very intriguing feature of above-threshold ionization or
high-order harmonic generation concerns the dependence of these phenomena on the
intensity of the driving field. The photo-electron or high-order harmonic peaks, as
functions of the driving-field intensity, present resonance-like
enhancements, such that a variation of a few percent in the external-field
strength may drive up the  spectral intensity by an order of magnitude. These enhancements have been observed experimentally by several
groups for above-threshold ionization (ATI) \cite{expati,cclos1}, and, recently,
for high-order harmonic generation (HHG) \cite{hhgres}. A concomitant effect in ATI is a variation of the contrast of the spectrum \cite{ago01}.

Early numerical observations that enhancements of ATI go hand in hand with
enhancements of HHG were reported in Ref.~\cite{PSK}. The existence of enhancements for 
{\it both} phenomena is not surprising, and is related to their common physical
origin. Indeed, HHG and ATI 
present very similar spectral features, which are
explained by similar physical pictures. These features are a wide energy
range with approximately equally strong harmonics or photo-electron peaks, 
known as ``the
plateau'', followed by a sharp decrease in the harmonic or photo-electron
signal, known as ``the cutoff''. HHG is described by
the so-called ``three-step model'', in which an electron is ionized through
tunneling or multiphoton ionization, is accelerated by the field and driven
back to its parent ion, where it recombines to the ground state, emitting its energy as one
harmonic photon \cite{tstep,lewen94}. A similar process is responsible for the plateau in
above-threshold ionization, with the main difference that, instead of
recombining with the parent ion, the electron is elastically rescattered off it 
\cite{hati}. In either case, the precise shape of the atomic potential is not very important, and it can be approximated, for instance, by a  zero-range potential. This 
approximation describes very well the spectral features near the
high-energy end of the plateau.

The resonance-like enhancements, however,  primarily occur in the first half
of the plateau, for which both the external driving field and the atomic
potential are expected to influence the harmonic or photo-electron emission 
\cite{cfmdbound}. In fact,  very different arguments have been put forward to explain these enhancements. Several studies attribute these features to  Rydberg
states that,  for an appropriate ponderomotive upshift, become multiphoton resonant with the ground state. Namely, a free electron in a laser field acquires a  field-dependent energy shift by the ponderomotive potential $U_{p}=e^{2}\left\langle
A^{2}(t)\right\rangle _{t}/2m$, where $A(t)$ is the vector potential of the laser field and 
$\left\langle ...\right\rangle _{t}$ denotes the average over a field cycle. Highly excited Rydberg states tend to undergo about the same shift as free electrons \cite{ago89}. The result of such a multiphoton resonance is either an
increase in ionization, or the electronic wave packet is trapped near its
parent ion for relatively long times, originating resonance-like structures
in the spectra \cite{muller1}. The mechanism is quite similar to the Freeman resonances \cite{freeman}, which dominate the low-energy ATI spectrum where rescattering plays no role. According to this physical picture, the
presence of high-lying Rydberg states is essential for the
existence of the enhancements.

An -- at least at first glance -- completely different view relates these
enhancements to channel closings that, by the same ponderomotive-upshift mechanism, may move into multiphoton resonance with the ground state. 
If, due to this shift, $N$ photons are no longer sufficient for the
electron to reach the continuum, one refers to the $N$-photon channel as
having closed. At an intensity corresponding to a channel closing, the electron is released in the
continuum with a vanishing drift momentum. In consequence, in the course of its oscillatory motion in the laser field, it will return many times to its
parent ion and upon each revisit have the opportunity to rescatter. 
Quantum mechanically, the corresponding probability amplitudes interfere, and a constructive interference manifests itself as an enhancement \cite{cclos1}.  Such an effect does not
require the existence of excited states or Rydberg states so that the atom can be modeled by a binding potential of zero range, which neither supports excited
bound states nor resonances in the continuum \cite{cclos1,Beck92,Sanp94}. 
The
zero-range potential affords the concept of ``quantum orbits'' which allows for an almost analytical approach to intense-laser--atom phenomena \cite{science}.
However, one
has to keep in mind that real atoms do have long-range-potential tails, so that channel closings are diffuse owing to the presence of the Rydberg series. In addition, they support various bound states,
whose influence on the enhancements is not entirely clear.

In this paper, we perform a systematic study of the influence of both the
laser field and the atomic potential on these enhancements, for
high-order harmonic generation, by means of simplified atomic models. We address the question of which one of the existing physical
interpretations is ultimately correct or whether both pictures are complementary
aspects of a more complete description. In particular, we investigate the
importance of the highly excited states in the process, and whether or not they are crucial for the feature in question.

\section{Truncated soft-core potentials}

We compute the harmonic spectra using the numerical solution of the
time-dependent Schr\"{o}dinger equation 
\begin{equation}
i{\frac{d}{dt}}|\psi (t)\rangle =\left[ {\frac{p\sp{2}}{2}}+V(x)-p\cdot
A(t)\right] |\psi (t)\rangle ,
\end{equation}
for a one-dimensional model atom \cite{footnote} initially in the ground state of a binding
potential $V(x)$ and subject to a laser pulse with the field $E(t)=-dA(t)/dt$. Atomic units are
used throughout. We consider a monochromatic laser field 
\begin{equation}
E(t)=E_{0}\sin \omega t,  \label{field}
\end{equation}
and the harmonic spectra are calculated from the dipole acceleration $\ddot{x%
}=\left\langle \psi (t)\right| -dV(x)/dx+E(t)\left| \psi (t)\right\rangle $ 
\cite{dipacc}. We take the smoothly truncated soft-core potential 
\begin{equation}
V(x)=\frac{-\beta }{\sqrt{\left( \frac{x}{\sigma }\right) ^{2}+1}}f(x),
\label{pote}
\end{equation}
with 
\begin{equation}
f(x)=\left\{ 
\begin{array}{c}
1\ (|x|<a_{0}), \\ 
\cos ^{7}[\pi \frac{|x|-a_{0}}{2(L-a_{0})}]\ (a_{0}<|x|<L), \\ 
0\ (|x|>L),
\end{array}
\right.   \label{trunc}
\end{equation}
so that $V(x)=0$ for $|x|>L$. We choose $a_{0}$ of the order of a few atomic units, and  $L$ of the order of  the electron
excursion amplitude, so that $L = r\alpha =rE_{0}/\omega ^{2}$ with the parameter $r$ of order unity. Setting  $f(x)=1$ in (\ref{pote}) gives the untruncated soft-core
potential. By an appropriate choice of the parameters $L$ and $a_{0}$, it is
possible to alter the highly excited bound states leaving the ground state and the low excited states practically unaffected. 

Let us assume that an atom, initially in the ground state with energy $\varepsilon_0$, is ionized by $N$
photons of frequency $\omega$, such that the electron reaches the continuum with the lowest energy possible, that is, with a drift momentum (outside the range of the binding potential) of zero. The energy of the $N$ photons must account for the binding energy and the kinetic energy of the oscillatory motion (the ponderomotive energy $U_{p}$), so that

\begin{equation}
|\varepsilon _{0}| +U_{p} =N\omega.  \label{channel}
\end{equation}
 For intensities slightly larger than specified by the condition (\ref{channel}), at least $N+1$ photons will be necessary for ionization, such that Eq.~(\ref{channel}) defines the 
$N$-photon channel-closing intensity. The intensities that solve  the channel-closing condition (\ref{channel}) form a comb whose 
teeth as a function of  $\eta =U_{p}/\omega $ are separated by unity. 
If there is an excited bound state with the (field-free) energy $\varepsilon_n$ and if this state undergoes the same ponderomotive upshift as the continuum \cite{ago89}, then  multiphoton resonance with the ground state occurs for intensities such that  Eq.~(\ref{channel}) is satisfied with $\varepsilon_0$ replaced by $\varepsilon_0 - \varepsilon_n$,

\begin{equation}
|\varepsilon _{0} - \varepsilon_n| +U_{p} =N\omega.  \label{excited}
\end{equation}

For a long-range potential [such as, in one dimension, our untruncated potential (\ref{pote})] the true continuum is preceded by the Rydberg series so that one may question the significance of the channel-closing condition (\ref{channel}) for any physical phenomenon. For a finite-range potential [such as the truncated potential (\ref{pote})] the Rydberg series is replaced by a finite sequence of bound states whose number decreases with decreasing $L$.

Below, we will seek to answer the following questions: Is the very existence of an enhancement contingent on the shape of the binding potential? If an enhancement exists, does it occur at a channel-closing intensity (\ref{channel}) for some $N$, or is related to a multiphoton resonance (\ref{excited}) with a certain excited state $|n\rangle$, or is it unrelated to either?  For a truncated 
potential, do enhancements occur both at the channel-closing intensities and at intensities where an excited state becomes multiphoton resonant? To what extent does the harmonic spectrum depend on the shape of the potential?

 For all examples presented in this paper, we
choose $\beta =2.1$ and $\sigma =0.2$ in the soft-core potential (\ref{pote}). In this case, the field-free energies  of the ground state and the  first four excited states are listed in Table~\ref{energies}  for the untruncated potential as well as for various truncations. The Table shows that the ground-state energy is virtually unaltered by the truncations we consider while the excited states are more and more affected. For the untruncated potential, the first excited states are followed by the Rydberg series. For all truncations considered, the states listed are the only ones that survived. By and large, for the parameter range used, the truncation
eliminates the Rydberg series, changes the energy of the third excited
state, and leaves the more deeply bound states unchanged. The shape of the effective
potential barrier $V_{{\rm eff}}=V(x)-xE(t)$ also remains very similar for
all cases.

\section{Results}

As a first step, we investigate a harmonic spectrum as a function of the laser intensity with regard to the existence of  enhancements for particular intensities. The frequency of the laser field is taken as $\omega =0.07600\ 
{\rm a.u.},$  such that Eq. (\ref{channel}) predicts channel-closing intensities  for near-integer values of $\eta =U_{p}/\omega.$ The other intensities that correspond to resonances with excited states are also listed in Table~\ref{energies}. For the frequency  and the intensity range considered in this paper, the Keldysh 
parameter $\gamma = \sqrt{|\varepsilon_0|/2U_p}$ lies within the interval $1.66<\gamma <0.86$, which is mostly 
within the multiphoton regime. 

\begin{figure}[tbp]
\begin{center}
\epsfig{file=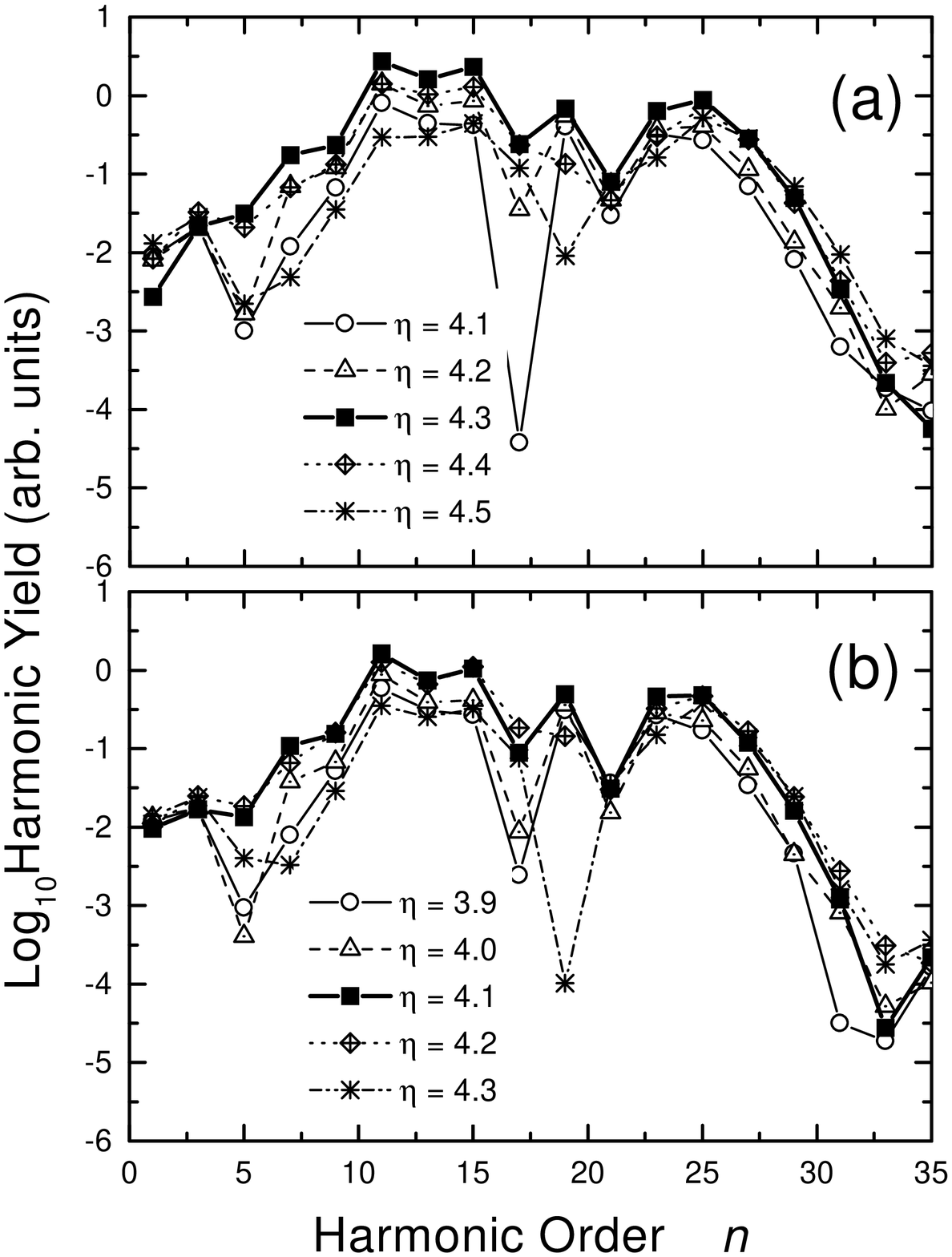,width=9.0cm,angle=0}
\end{center}
\caption{Harmonic spectra for $\omega = 0.076$ a.u., $\varepsilon_0 = 0.7566$ a.u., and several values of $\eta =U_{p}/\omega ,$ for the untruncated (a) and
truncated (b) soft-core potential, with $L\simeq 31.78\ {\rm a.u}.$ and $a_{0}=3\ 
{\rm a.u}$. The parameter $L$ was
chosen as twice as the electron excursion amplitude $\alpha$ for $\eta=4.8$. 
The
harmonic intensities are connected by lines to guide the eye. The thick lines correspond
to the field intensity for which the enhancement is maximal.}
\label{harm}
\end{figure}

 Figure \ref{harm} displays two harmonic spectra, computed for the   
untruncated (a) and a truncated (b) soft-core potential. Both spectra are very 
similar, with 
pronounced enhancements near the 13th harmonic.  For the untruncated and 
truncated cases, we observe maximal enhancements 
 at $\eta =4.3$ and  $\eta =4.1$, respectively. These values do not coincide 
with the channel-closing intensity $\eta =4.04$ (modulo any integer) predicted by Eq.~(\ref{channel}). This shows a clear influence of the binding potential on the
absolute intensity for which these features occur. However, comparing parts (a) and (b) of Fig. \ref{harm} we notice a very remarkable fact: the spectra of the untruncated (a) and the truncated (b) potential are almost identical provided we compare  a spectrum at the intensity $\eta$ for the former with a spectrum at the intensity $\eta -0.2$ for the latter. This holds for all intensities considered. Table~\ref{energies} shows that the truncated potential no longer 
supports the Rydberg series and its excited state $|3\rangle$ is very close to the continuum. We conclude that the Rydberg series has no visible impact on the shape of the harmonic spectrum. Checking again Table~\ref{energies} we further conclude that the intensities where the enhancements are maximal are compatible with Eq.~(\ref{excited}) with $n=3$.

\begin{figure}[tbp]
\begin{center}
\epsfig{file=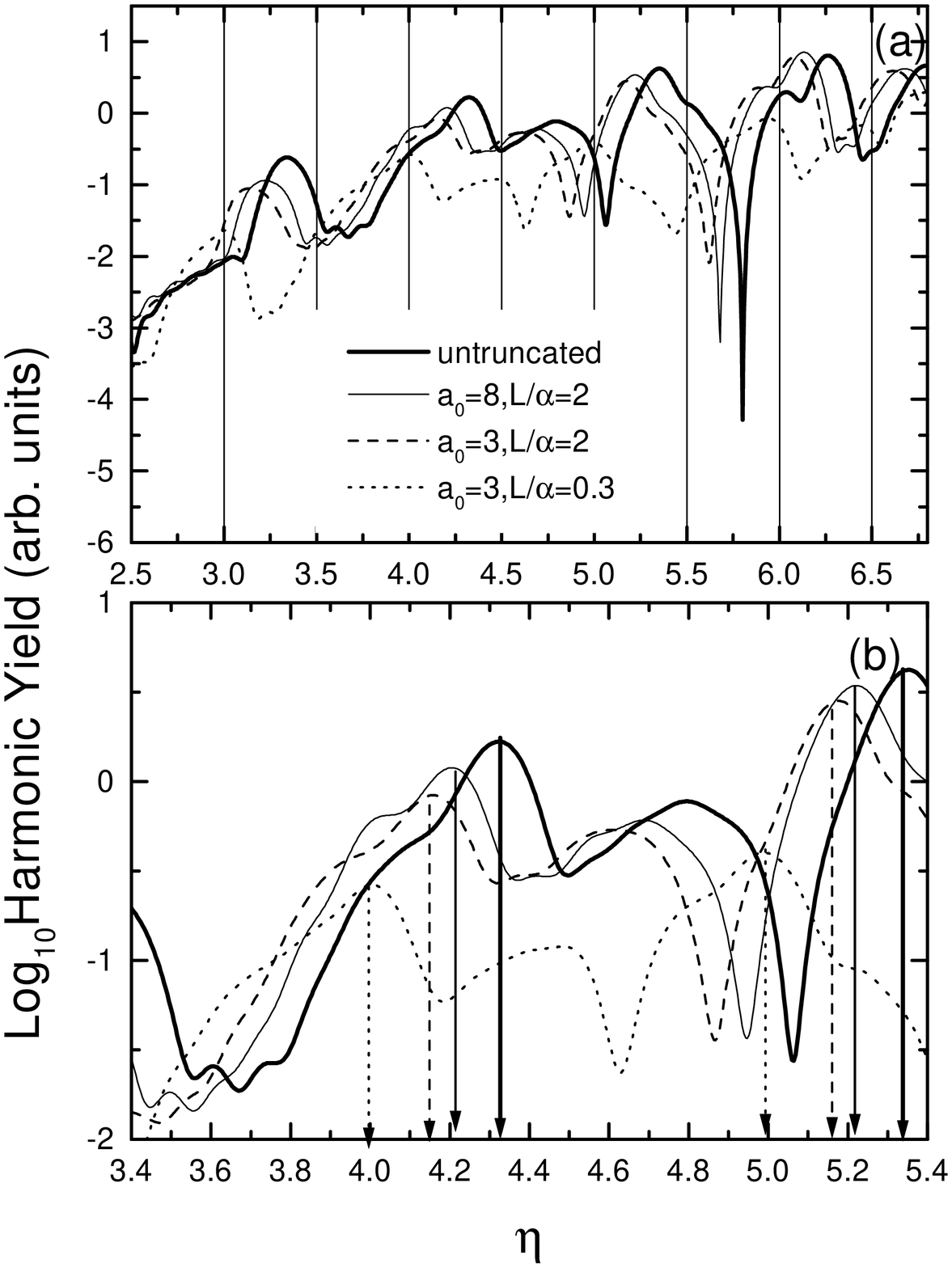,width=9.0cm,angle=0}
\end{center}
\caption{Intensity of the 13th harmonic as function of $\eta =U_{p}/\omega .$
Part (a): Comparison
between the untruncated and various truncated soft-core potentials; the two values of the parameter $L$, $L=31.78$ a.u. and $L=4.77$ a.u., correspond to $2\alpha(\eta=4.8)$ and $0.3\alpha(\eta=4.8)$, respectively. The binding energies for the various truncated potentials are given in table \ref{energies}. Part (b): Enlargement of part (a),
for $3.4<\eta <5.4.$ The values of this parameter, for which the enhancements
 are maximal, are marked by arrows.}
\label{harm13}
\end{figure}

Further support for these conclusions comes from Fig. \ref{harm13}, where the yield of the 13th
harmonic is plotted as a function of the scaled intensity $\eta = U_p/\omega$. In this figure,
we also investigate the effect of the truncation in more detail, for a wider
range of intensities and truncating parameters. The most striking feature is
that the main effect of the various truncations is a rigid horizontal shift of the yield-versus-intensity curve. Remarkably, this statement includes the pronounced dips which are due to quantum interference between different quantum 
orbits \cite{lewen94,science}. While this may be plausible for the large values of the truncation parameter  $(L = 2\alpha)$ where the truncation mostly affects the potential outside of the classical electronic excursion, it is quite surprising for the small value $L = 0.3 \alpha$.  A similar shift pattern is observed for all low plateau harmonics (not shown).

Next, we investigate the quantitative amount of the afore-mentioned shift.  Comparison with Table~\ref{energies} shows that all of the enhancement peaks present 
in Fig.~\ref{harm13} are compatible with the intensities predicted by  Eq.~(\ref{excited}) for $n=3$. As long as the state $|3\rangle $ exists as a bound state, we conclude that the enhancement is due to a multiphoton resonance with this state, upshifted by the whole amount of the ponderomotive energy. 
However, the enhancement still exists (near integer $\eta$), though quantitatively smaller,  in the case where the truncation (for $L=0.3\alpha$) has eliminated the third excited state. Here we have encountered a case of a pure channel-closing enhancement which is due to a multiphoton resonance with the ponderomotively upshifted continuum threshold.

\begin{figure}[tbp]
\begin{center}
\epsfig{file=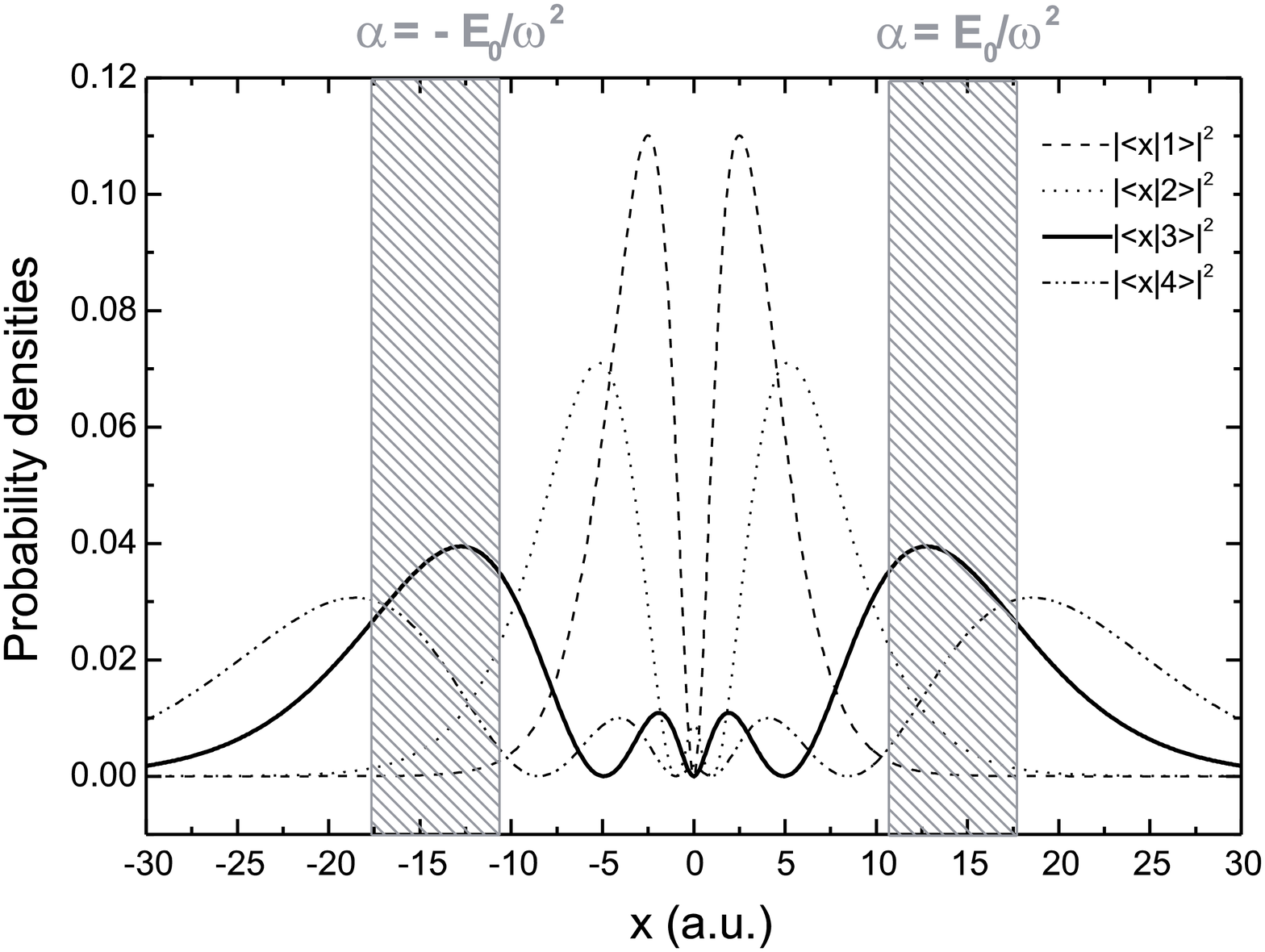,width=9.0cm,angle=0}

\end{center}
\caption{The probability distributions $|\langle x|n\rangle |^2$ for the first four excited states of the untruncated potential (\ref{pote}). The range of excursion amplitudes $\alpha = E_0/\omega^2 = 2\sqrt{\eta/\omega}$ that is covered in Fig.~\ref{harm13} (for $2.3<\eta <6.3$) is shaded. Truncation has a minor effect for $n=1$ and 2. For $n=3$ this effect is becoming noticeable, and for $n=4$ it is strong. All higher excited states are eliminated by the truncations considered.}
\label{xmax} 
\end{figure}

Why does a multiphoton resonance with the third excited state (as long as it exists) lead to a pronounced resonance in the spectrum while a multiphoton resonance with other excited states does not? We can answer this question by inspecting the wave functions $\langle x|n\rangle$ of the excited states. Please, notice that our calculations are still within the multiphoton regime of ionization. 
In this case, one expects that multiphoton resonance with an excited bound state is particularly relevant if the wave function of the latter is concentrated near the turning points of the wiggling motion of a classical electron with a drift momentum of zero, i.e., when $|\langle x|n\rangle |^2$ has its maxima near $|x| \approx \alpha = E_0/\omega^2$. Figure~\ref{xmax} shows that indeed this is well satisfied by the state $|3\rangle $.

\begin{figure}[tbp]
\begin{center}
\epsfig{file=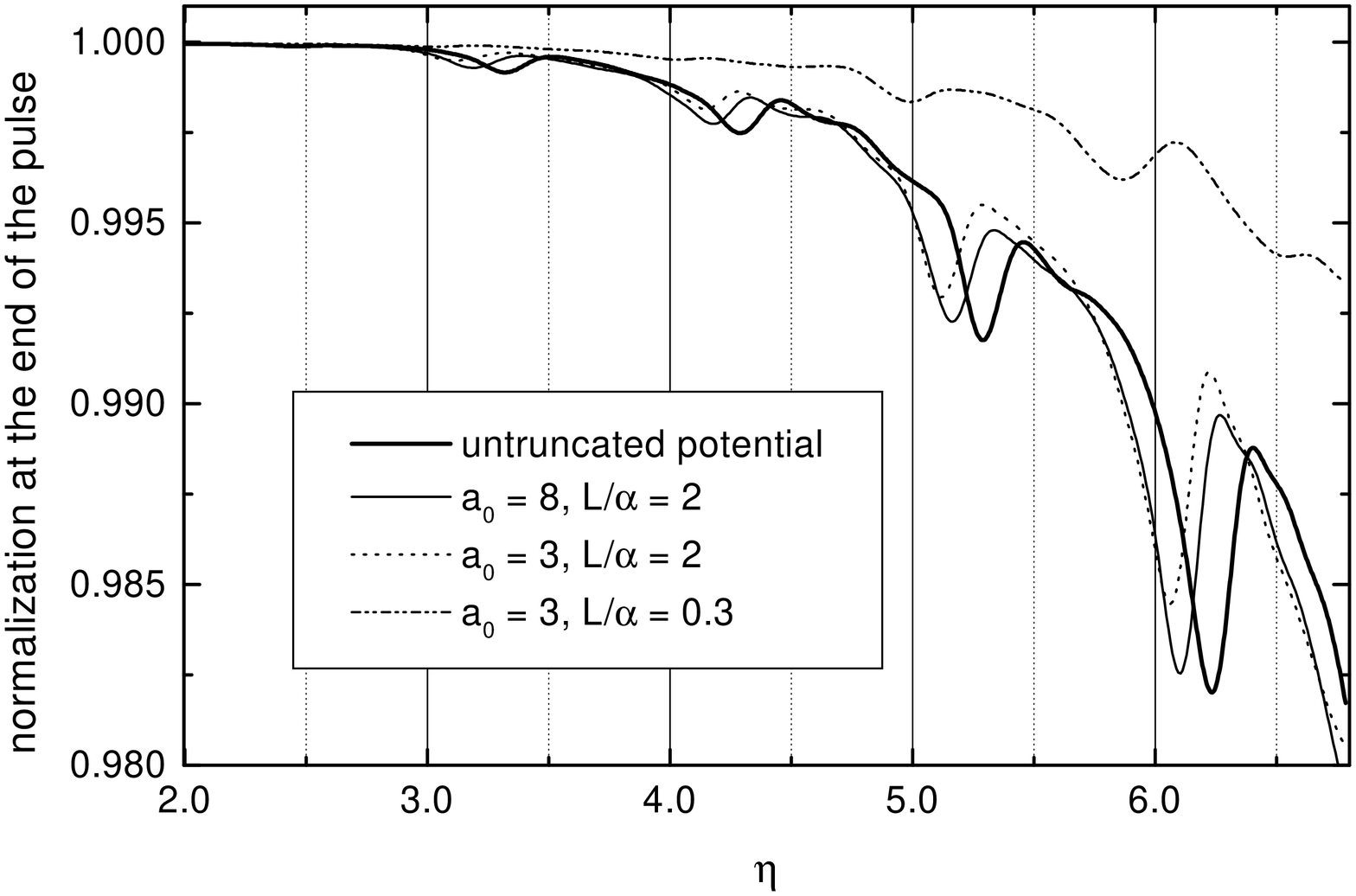,width=6.4cm,angle=270}
\end{center}
\caption{Normalization of the time-dependent wave function at the end of the
pulse as a function of $\eta=U_p/\omega$, for the truncation parameters of
Table~\ref{energies}. The pulse length is eight cycles plus a linear turn-on of two cycles.}
\label{norm}
\end{figure}

It is not surprising that ionization is affected by the resonances in a similar fashion as HHG. In Fig.~\ref{norm} we plot the normalization of
the time-dependent wave function at the end of the pulse, as a 
function of $\eta$ for the truncation parameters used in Fig.~\ref{harm13}. 
This normalization is smaller than unity, since part of the
wave function is absorbed by a mask 
function, used in our computations to eliminate spurious reflection effects.
It is a good measure of irreversible ionization, since the mask function
is located at about ten times  the electron excursion amplitude. In the figure, there are clear dips
slightly preceding the intensities, 
for which the channel closings occur. 
The particular case, for which 
$\left| 3\right\rangle$ is absent, agrees with the results in \cite{Sanp94} 
for a zero-range potential.

There are still features in the enhancements, such
as their strength and harmonic range, which depend on the atomic binding
potential and are not yet completely understood. In fact, studies performed
by us for short- and long-range potentials of various shapes indicate that, even though the
enhancements are a generic feature being present for all cases, their
strength, regularity and harmonic range depend on the potential in
question. An interesting feature is the fact that the enhancements are
weaker for short-range potentials. Since the tail of the potential plays no role in
this respect, the effective potential barrier may influence these features.

\section{Discussion and conclusions}

The picture that emerges from these calculations is this: high-order harmonic generation can be strongly enhanced by a multiphoton resonance with a ponderomotively upshifted  excited bound state close to the continuum threshold. In the cases that we investigated the relevant (field-free) bound state $|n\rangle $ was the state where $|x_{n,{\rm max}}| \approx \alpha \equiv E_0/\omega^2$ with $x_{n,{\rm max}}$ the value of $x$ that renders $|\langle x|n\rangle |^2$ maximal. Bound states higher than this do not lead to noticeable enhancements, 
nor do the existence or nonexistence of a Rydberg series have an effect, neither on the enhancements nor on the entire harmonic spectrum. In all cases considered, the relevant bound state was $|n=3\rangle $. When this state was eliminated by sufficient truncation of the potential, the role of the resonant bound state was taken over by the ponderomotively upshifted continuum threshold. Hence, in all 
cases, the resonant intensities are well described by Eq.~(\ref{channel}), provided we employ an effective continuum threshold that is given by $\tilde\varepsilon \equiv \varepsilon_0 - \varepsilon_{\tilde n}$ where $\varepsilon_{\tilde n}$ is the energy of the crucial bound state if such a state exists and zero 
otherwise. Remarkably, if the value of $\varepsilon_{\tilde n}$ changes owing to a change of the potential, the effect on the harmonic spectrum is largely a horizontal shift of the yield-versus-intensity curve of the various harmonics.

This picture allows one, for the purposes of HHG, to model the atom by a short-range or zero-range potential using a binding energy that is adapted to the energy difference between the ground state and the relevant excited state of the real atom that is supposed to be modeled. In this case, excited states need not be considered and the enhancement can be attributed to a multiphoton 
resonance with the continuum threshold and the generation -- in the three-step model -- of an electron with a vanishing drift momentum. The picture of an effective continuum threshold is also intuitively appealing: in the presence of a strong laser pulse, the highly excited states acquire finite widths owing to ionization and the finite pulse duration, so that the electron moves in a quasi-continuum. 

In the case that we investigated $(a_0=3,\ L=0.3\alpha)$, cf. Fig.~\ref{harm13}, the enhancement as well as the general harmonic yield were markedly reduced compared with the untruncated potential. This is not so for a comparison of a realistic single-active-electron binding potential to a (regularized) zero-range potential in three spatial dimensions, neither for HHG nor for ATI \cite{cclos2}.

In conclusion, we have confirmed the significance of resonant enhancements for high-order harmonic generation. The resonances occur, when an appropriate highly excited state or, in the absence of such a state, the continuum threshold are 
ponderomotively upshifted so that they become multiphoton resonant  with the 
ground state. The mechanism is similar to high-order above-threshold ionization. In the latter case, remarkably, the multiphoton resonance continues to be significant while the ionization process is already deeply in the tunneling regime \cite{muhelium}. For high-order harmonic generation, this remains to be investigated.

\section*{Acknowledgments}
We enjoyed discussions with M. Kleber, H. G. Muller, and G. G. Paulus. This work was supported in part by Deutsche Forschungsgemeinschaft.

\onecolumn
\begin{table}
\caption{ Energies $\varepsilon_n$ of the field-free ground state and the first four excited states for the untruncated potential (\ref{pote})  as well as several truncated versions thereof, as considered in Fig.~\ref{harm13}. No entry means that the state is no longer bound. Multiphoton resonances with the respective state occur for intensities $\eta_n = {\rm integer} + \Delta \eta_n$.}
\label{energies}
 \begin{tabular}{ccccccccc}
\mbox{} & \multicolumn{2}{c}{untruncated} & \multicolumn{2}{c}{$a_0=8\ \ L=31.78$} & \multicolumn{2}{c}{$a_0=3\ \ L=31.78$}& \multicolumn{2}{c}{$a_0=3\ \ L=4.77$}\\
n & $\varepsilon_n$ & $\Delta \eta_n$ & $\varepsilon_n$ & $\Delta \eta_n$ & $\varepsilon_n$ & $\Delta \eta_n$ & $\varepsilon_n$ & $\Delta \eta_n$ \\
\hline \vspace*{-0.5ex}\\
0 & 0.7566 & 0.04 & 0.7566 & 0.04 & 0.7566 & 0.04 & 0.7565 & 0.05 \\
1 & 0.0846 & 0.16 & 0.0845 & 0.16 & 0.0825 & 0.13 & 0.0524 & 0.74 \\
2 & 0.0465 & 0.67 & 0.0453 & 0.64 & 0.0389 & 0.55 &  - & - \\
3 & 0.0216 & 0.33 & 0.0124 & 0.21 & 0.0036 & 0.09 & - &- \\
4 & 0.0156 & 0.25 & 0.0014 & 0.06 & - &- &- &-
\end{tabular} 
\end{table}

\end{document}